*IEEE* Access

Multidisciplinary : Rapid Review : Open Access Journal

# A review on the mobile applications developed for COVID-19: An exploratory analysis

MUHAMMAD NAZRUL ISLAM [ID]1, IYOLITA ISLAM [ID]1, KAZI MD. MUNIM [ID]1, AND A.K.M. NAJMUL ISLAM [ID]2, 3

[1]Department of Computer Science and Engineering, Military Institute of Science and Technology (MIST), Mirpur Cantonment, Dhaka-1216, Bangladesh
[2]LUT School of Engineering Science, LUT University, 53850 Lappeenranta, Finland
[3]Department of Future Technologies, University of Turku, Finland

Corresponding author: Muhammad Nazrul Islam (e-mail: nazrul@cse.mist.ac.bd)

**ABSTRACT** The objective of this research is to explore the existing mobile applications developed for the COVID-19 pandemic. To obtain this research objective, firstly the related applications were selected through the systematic search technique in the popular application stores. Secondly, data related to the app objectives, functionalities provided by the app, user ratings, and user reviews were extracted. Thirdly, the extracted data were analyzed through the affinity diagram, noticing-collecting-thinking, and descriptive analysis. As outcomes, the review provides a state-of-the-art view of mobile apps developed for COVID-19 by revealing nine functionalities or features. It revealed ten factors related to information systems design characteristics that can guide future app design. The review outcome highlights the need for new development and further refinement of the existing applications considering not only the revealed objectives and their associated functionalities, but also revealed design characteristics such as reliability, performance, usefulness, supportive, security, privacy, flexibility, responsiveness, ease of use, and cultural sensitivity.

## I. INTRODUCTION

CORONAVIRUSES (CoVs) are a family of viruses that can infect human beings as well as different species of animals like cattle, cats, and bats [1]. These viruses can cause illness varying from the very mild cold to severe health conditions including death. A pneumonia, referred to as Severe Acute Respiratory Syndrome (SARS-CoV), was first recognized in China [2] with an overall death rate of around .3 to 6% [3]. Like SARS-CoV, Middle East Respiratory Syndrome (MERS-CoV) is also a member of CoVs family. And, the 2019 coronavirus (COVID-19) is a new betacoronavirus, which has infected humans recently. Common symptoms of COVID-19 virus include fever, cough, and breathing difficulties, among others. In severe cases, the infection can cause pneumonia and even death. People with previous diseases related to heart, kidney, lungs, and diabetes are more likely to develop serious illness from COVID-19. After China, COVID-19 has been spread in more than 210 countries and territories [4]. On March 11, 2020, the World Health Organization (WHO) has declared COVID-19 as the global pandemic [5].

The immediate risk of being infected by COVID-19 is increasing day by day as the outbreak expands. Since there had been limited or delayed precautionary responses against the new virus, it had spread rapidly from person-to-person globally. It has been suggested that further COVID-19 spread can be controlled by taking proper and urgent precautions and raising health awareness among people [6]. Several International organizations like WHO [7], Centers for Disease Control and Prevention (CDC) [8] have suggested necessary plans to deal with the pandemic. Countries have taken steps such as providing health safety guidelines, imposing travel restrictions, organizing enough quarantine and hospital facilities for infected people, and even complete lock-down [9] [10] [11] [12]. To prevent virus infection some suggestions include: frequent hand washing, avoiding crowded places, covering mouth and nose, eating fully cooked meat and eggs, and avoiding close contacts with anyone coughing and sneezing, among others [1].

Some information and communication technology (ICT) based initiatives have been taken for raising awareness for preventing the pandemic caused by COVID-19. These in-clude mobile applications, online dashboards to provide live updates, websites to provide remote assistance, web portals to provide useful information related to COVID-19, and the development of electronic gadgets to detect or scan the symptoms of COVID-19, among others [13] [14] [15].

At present, more than 45% people in the world use smartphones [16]. Indeed, smartphones and Internet penetration in the developed world is high. At the same time, recent studies show that the uses of smartphones and the Internet have also been noticeably high in many developing countries [17]. As an example, a study conducted by Islam et al. [18] found that around 234 mobile health (mHealth) applications were developed explicitly for Bangladesh. Developing countries have relatively weak healthcare facilities. Therefore, there is a possibility that COVID-19 may hit hard in these countries. Mobile apps can support interactivity, visual and auditory content, real-time data collection, as well as links to social functionalities [19] [20]. Considering the widespread adoption of smartphones in the developing and developed countries, there is a growing opportunity for the effective uses of mobile applications for preventing, caring and controlling the pandemic caused by COVID-19.

A number of review studies were conducted focusing on the mobile health applications developed for cardiovascular disorder [21], depression [22], human immunodeficiency virus [23] [24], cancer [25], schizophrenia [26], monitoring physical activity [27], and managing bipolar disorder [28], among others. However, up to date, no study has assessed the available apps related to the COVID-19 pandemic. Indeed, as the COVID-19 pandemic is a newly developed phenomenon and therefore, available apps in this area are still limited and more efforts in app development will follow in the future. However, at the same time, we believe that as it has become a worldwide pandemic, research needs to investigate what apps

**TABLE 1.** Summary of the reviewed apps related to COVID-19

| Ser. | Name | Star rating | First Released | Version | Cost (US $) | Platform | Privacy Policy | Country-context | Language |
|---|---|---|---|---|---|---|---|---|---|
| A1 | Covid-19 | 3.8 by 15 users | Mar 2020 | 1.0.4 | Free | Android | Yes | Global | English |
| A2 | HealthLynked COVID-19 Tracker | 4.2 by 136 users | Feb 2020 | 2.2.1 | Free | Android | Yes | Global | English |
| | HealthLynked COVID-19 Tracker | 4.6 by 18.1K users | Feb 2020 | 1.0.26 | Free | iOS | Yes | Global | English |
| A3 | Coronavirus COVID-19 Tracker | No rating yet | Mar 2020 | 1.0.5 | Free | Android | Yes | Global | English |
| A4 | COVID19 Monitor | 4.1 by 8 users | Mar 2020 | 1.0.3 | Free | Android | Yes | Global | English |
| A5 | World Virus Watch | 4.1 by 47 users | Mar 2020 | 1.3.1 | Free | Android | Yes | Global | English |
| A6 | Coronavírus - SUS | 3.9 by 6.6K users | Feb 2020 | 1.0.8 | Free | Android | Yes | Brazil | Portuguese |
| | Coronavírus - SUS | 3.6 by 85 users | Feb 2020 | 1.1.3 | Free | iOS | Yes | Brazil | Portuguese |
| A7 | COVID Live Tracker - Corona Virus Pocket Guide | 4.5 by 15 users | Mar 2020 | 1.5 | Free | Android | Yes | Global | English |
| A8 | STOP COVID-19 EN FRANCE | No rating yet | Feb 2020 | 1 | Free | Android | Yes | France | French |
| A9 | Suc khoe Viet Nam | 4.5 by 2K users | Feb 2020 | 1.0.1 | Free | Android | Yes | Vietnam | Vientamese |
| A10 | COVID-19 | 3.2 by 13 users | Mar 2020 | 2.1 | Free | iOS | Yes | Vietnam | Vientamese |
| A11 | COVA Punjab | 4.6 by 51 users | Mar 2020 | 1.0.0 | Free | Android | Yes | India | English |
| | COVA Punjab | 2 by 4 users | Mar 2020 | 1.0.3 | Free | iOS | Yes | India | English |
| A12 | nCovi MobiFone | 2.7 by 40 users | Mar 2020 | 1.2 | Free | iOS | Yes | Vietnam | Vientamese |
| A13 | Helponymous: Corona-Virus Chat | 4.8 by 60 users | Feb 2020 | 2.5 | Free | iOS | Yes | Global | English |
| A14 | COVID19_GR | 5 by 1 user | Feb 2020 | 1 | Free | Android | Yes | Greece | Greek |
| A15 | COVID19 - Live Map of Coronavirus | No rating yet | Feb 2020 | Unavailable | Free | Windows | Yes | Global | English |
| A16 | CoronaVirus updates - Live Tracker & Map | 5 by 2 users | Feb 2020 | Unavailable | Free | Windows | Yes | Global | English |
| A17 | COVID-19 Sounds | 4.4 by 16 users | Apr 2020 | 1.1.12 | Free | Android | Yes | Global | English |
| A18 | CoronaBD | 4.3 by 136 users | Apr 2020 | 1.1.1 | Free | Android | Yes | Bangladesh | Bangla |
| A19 | CMED Agent | 4.5 by 76 users | Apr 2020 | 2.1.0.1 | Free | Android | Yes | Bangladesh | Bangla |
| A20 | COVID Symptom Tracker | 4.7 by 66,181 users | Mar 2020 | 0.12 | Free | Android | Yes | UK | English |
| | COVID Symptom Tracker | 4.7 by 98.9K users | Mar 2020 | 1.7 | Free | iOS | Yes | UK | English |
| A21 | Canada COVID-19 | 4.6 by 331 users | Mar 2020 | 2.5.0 | Free | Android | Yes | Canada | English |
| | Canada COVID-19 | 4.5 by 370 users | Mar 2020 | 2.5 | Free | iOS | Yes | Canada | English |
| A22 | STOP COVID19 CAT | 3.3 by 1663 users | Mar 2020 | 1.0.2 | Free | Android | Yes | Catalonia, Spain | Catalan |
| | STOP COVID19 CAT | 4.2 by 5 users | Mar 2020 | 1.0.2 | Free | iOS | Yes | Catalonia, Spain | Catalan |
| A23 | Apple COVID-19 | 4.3 by 2.1k users | Mar 2020 | 2.2 | Free | iOS | Yes | Global | English |
| A24 | COVID-19! | 3.3 by 34 users | Mar 2020 | 0.9.4 | Free | Android | Yes | Global | English |
| | COVID-19! | 4.1 by 54 users | Mar 2020 | 0.9.7 | Free | iOS | Yes | Global | English |
| A25 | Coronavirus Australia | 3.3 by 1358 users | Mar 2020 | 1.0 | Free | Android | Yes | Australia | English |
| A1, A2, A3,....,A25 : The number of the mobile applications. | | | | | | | | | |

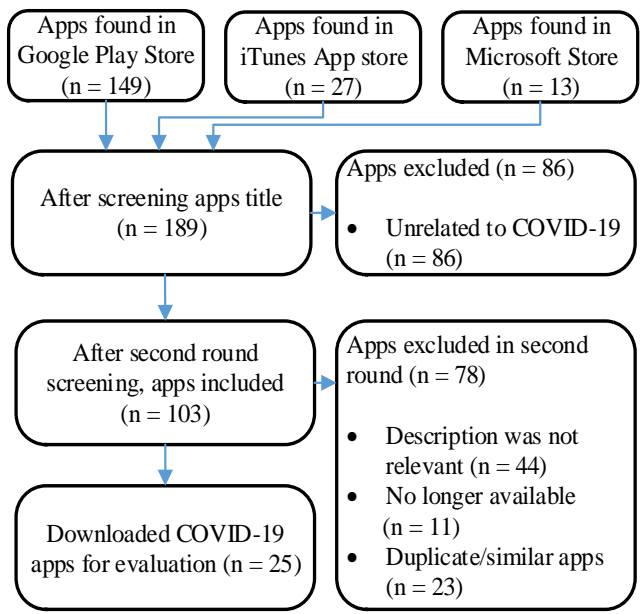

**FIGURE 1.** App inclusion and exclusion process flowchart

have been developed and make a synthesis to guide future service development.

In order to address the above-mentioned research gap, this study conducts a thorough review of the existing mobile apps that have been developed focusing on COVID-19. Our objectives are to (i) identify existing apps related to the COVID-19 in the popular app stores; (ii) investigate the apps in terms of their objectives and associated features or functionalities; (iii) explore the platform (operating system) of the apps, country-context for which the app was developed, and the language of the app; and (iv) identify important design characteristics based on the positive and negative feedback from users.

## II. METHODS
Reviewing the existing apps developed for COVID-19 was conducted by a systematic app review. The study was carried out following the steps discussed in the following subsections.

### A. SYSTEMATIC SEARCH CRITERIA AND SELECTION
A systematic review of mobile applications across three major mobile app stores: Android Google Play, Apple Store, and Microsoft Store was conducted on 30th April 2020. The search keywords used to find out the related applications were: "coronavirus," "COVID 19," "corona," "corona outbreak," "corona pandemic," "corona and symptom monitoring," "corona and self-care," "COVID 19 and symptom monitoring," "corona and COVID 19," "corona and self-isolation," "corona and quarantine," and "self-care and COVID 19." Each term was searched in each of the three app stores listed.

App titles, screenshots of the apps, and app descriptions were considered for the preliminary screening. The second round of exclusion criteria focused on removing the duplicate apps that were found in multiple stores or from multiple search terms or the apps that were no longer available (some of the apps were removed from the app stores). Two members of this research team meticulously reviewed the apps in each round of exclusion. Finally, a total of 25 apps were selected, downloaded, reviewed, and evaluated. The description of the selected applications is provided in Table 1. The related app screening, inclusion/exclusion, and selection process are shown in Figure 1.

### B. EXTRACTING THE STUDY DATA
To attain the research goal, we extracted data related to the objective of the apps, functionalities provided by the apps, the platform (operating system) of the selected apps, country-context, language of the apps, user ratings, first release, and user comments. The objectives and features of the apps were extracted both from the app description and by the experimental use of the apps. For extracting the features, we first went through the app description and kept notes of the stated features. In some cases, when the app language was not English, we used Google translator. After that, each app was installed and explored to re-check (verify) the extracted features and update the list of extracted features as observed from the experimental use of the app. We ensured that the features extracted against each app from the description must be present in the actual app through the experimental use of the app. Our approach helped us to explore the stated features as well as other features (if any) from our experimental use of the app.

### C. DATA ANALYSIS
An affinity diagram [29] approach was used to analyze the data to find out the objectives of the apps. A *noticing-collecting-thinking* approach [30] was used to analyze the user comments ($n = 1574$) of the selected applications to identify the design characteristics of the apps. The other data including the review scores, development platform, application languages, and the country-context were analyzed using descriptive statistics and data synthesizing. We note that our approach of data collection from online sources is widely used in different disciplines [31]–[33], and known as netnography [34].

Each of the selected apps was investigated to extract their functionalities or features. Each author of this article separately participated in exploring the app functionalities and then in grouping (mapping or clustering) process through an affinity diagram [29]. Firstly, each author extracts the functionalities and assigned names to the functionalities. After that they met together to compare and verify the extracted functionalities for each application. The naming of the extracted functionalities, conflicting functionality names, and disagreements were resolved by discussion through consensus to finalize the functionalities of the selected apps. Secondly, each researcher separately categorized the extracted functionalities into different groups based on their similarities. Next, an intuitive name was given to each group (which are considered as the objective of the application) and then drew the affinity diagram. Finally, the four affinity diagrams (by four authors) were analyzed together to review the grouping or clustering. Thereafter, we came up with one affinity diagram with the final clustering.

Since the user review data are qualitative in nature and not all the reviews were related to the design issues, we have followed the *noticing-collecting-thinking* [30] method to analyze each of the retrieved review comments. An example of analyzing a review comment through the noticing-collecting-thinking approach is discussed here. Firstly, at the noticing level, we read a review comment to notice an interesting or relevant aspect that may relate to the design of the application and coded it accordingly. For example, the following user review comment was noticed, which was relevant to the application design: *"I did not like that you have to sign up with your phone number before you can enter the app. I am concerned about digital privacy and this is just not okay, especially in this crucial time when information is crucial."* The extracted code of this review comment was 'concerned about digital privacy'. Secondly, in the collecting level, all the related codes were grouped and given a self-expressive or self-descriptive group name. The 'concerned about digital privacy' code was grouped under the category of 'Privacy'. Finally, in the thinking stage, we examined the things (e.g., all codes related to 'privacy') that we had collected to verify whether the things make some type of sense (e.g., a design characteristic or recommendation) out of each collection. Furthermore, we checked whether any relationship exists within a collection and across the collection. While collecting and thinking, a review comment was coded as either positive or negative based on its referential or indicative meaning. For example, if any user review expresses the difficulties of using a feature or reveals dissatisfaction with the app (e.g., providing unauthentic and outdated information), then we have considered such a statement as a negative comment. The example quote stated above expresses dissatisfaction of the user as he/she was asked to sign up using a phone number and therefore, was concerned about digital privacy. Thus, the above comment was coded as a negative comment.

The review comments were independently analyzed by two researchers. The comments were meticulously read to notice, and code as we discussed above. The assigned key-words or codes were then collected and categorized to represent common themes. After that, both researchers thought about their coding to refine, verify and update the code names, categorizations, relations among the categorizations. Finally, all authors met to compare the coding and classifications. The inter-coder agreement [35] was 0.91. The disagreements were resolved by discussion through consensus. Using an iterative process, all researchers read, sorted, reread, and recombined the data until consensus was achieved.

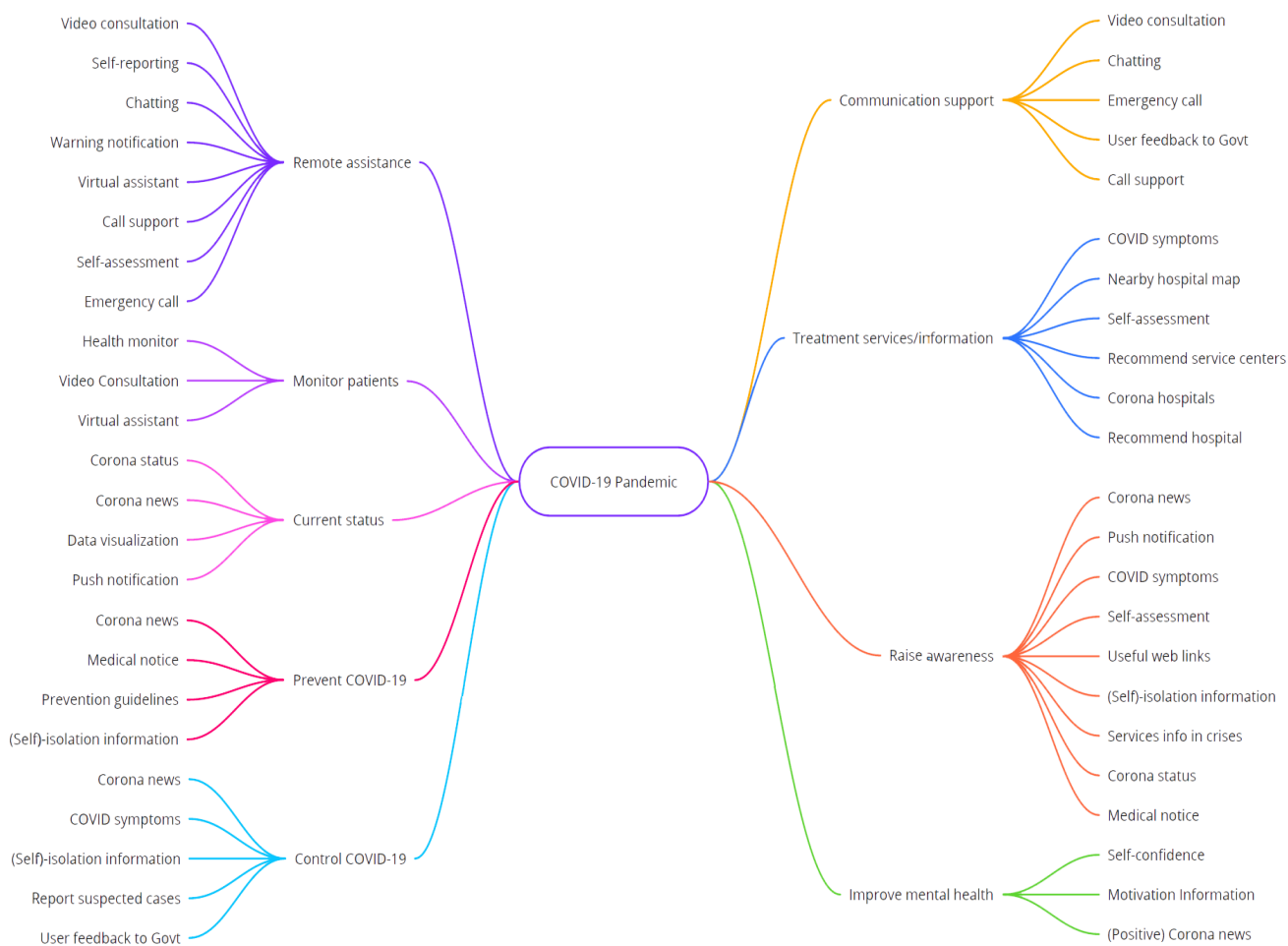

**FIGURE 2.** Mind map of the features and objectives of the apps

## III. RESULTS

This section discusses the findings of reviewing the selected mobile applications.

### A. APP OBJECTIVE AND FEATURES

As an outcome, a total of 26 functionalities were found from the 25 mobile applications. The extracted functionalities for each mobile application are shown in Table 3 of Appendix A. Through the affinity diagram, a total of nine objectives were found for which the mobile applications were developed. The extracted functionalities and their mapping to the objectives of the mobile applications are presented in Figure 2. The objectives of the mobile applications were also mapped to the selected mobile apps and shown in Table 4 of Appendix A. The results (see Table 4) show that most of the applications were developed for the multiple purposes but there was no single application found that addressed all objectives or purposes. The number of applications that offered a specific objective is shown in Figure 3. The revealed purposes of the mobile applications for the COVID-19 pandemic are briefly discussed below.

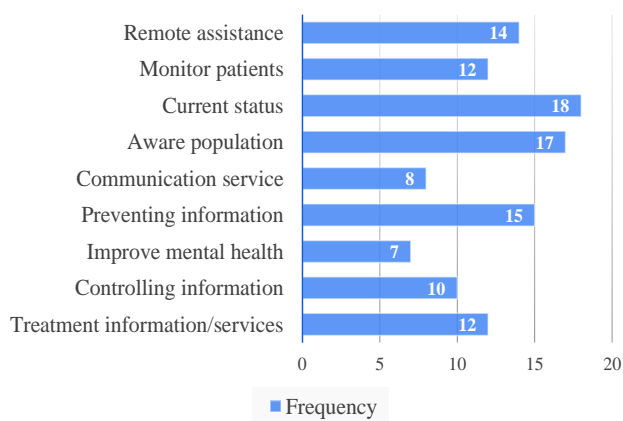

**FIGURE 3.** The objectives of the mobile applications

1) Remote assistance - About 56 % of the applications (i.e., 14 out of 25 applications) provide remote assistance to users or patients. The functionalities to achieve this objective are remote consultation through video communication, audio call, instant messaging, and emergency call support. Patients may conduct self-assessment to check if their symptoms are related to COVID-19. Some applications support self-reporting of COVID-19 infection as well as receiving the virtual medical assistance. A few apps also provide warning notifications on a regular basis related to new affected geographical regions and awareness information about COVID-19.

2) Monitor patients - Around half of the applications (i.e., 12 out of 25 applications) include features for monitoring patients by observing the users' health parameters such as temperature, heart rate, oxygen, and blood pressure using Bluetooth certified medical devices. The patients can consult with doctors through live video and receive medical advice. Some apps provide virtual assistance through interacting with people in natural language and voice.

3) Current status - Most of the applications (i.e., 18 out of 25 applications) provide the current status and statistics from a local (specific-country context) and a global perspective through data-sheets and data visualizations. Such status includes the number of confirmed infected people, recovery, death, incubation period, etc. with respect to the time and geographical location. Some apps provide the prediction data through different kinds of mathematical analyses; and some other apps provide recent corona virus related news and push notifications for any new suspected case, suspected community/regions, death, and recovery.

4) Prevent COVID-19 - More than half of the selected applications (i.e., 15 out of 25 applications) provide services for the prevention of COVID-19 by providing updated COVID-19 related news about newly suspected community/regions. Especially, these applications provide the preventive guidelines suggested by the ministry of health, WHO, CDC, or other authentic body. Some apps focus to provide the necessary information for staying at self-isolation.

5) Control COVID-19 - Around 40% of the applications (i.e., 10 out of 25 applications) target to control the spread of COVID-19 by providing instructions for self-isolation, facilitating to report the suspected cases to the health authority, nearest hospital, or to the government; and conveying the users feedback to the government bodies to take necessary initiative to control COVID-19. Some other applications provide information related to COVID-19 symptoms and keep people updated with the situation by providing updated news regularly, which also assists end-users to control COVID-19.

6) Communication support - Around one-third of the applications (i.e., 8 out of 25 applications) include services to provide communication support using live chat or video consultation, instant messaging, regular or emergency phone calls to hot-lines/dedicated help numbers. A few applications provide interaction/communication facilities between the government bodies and people. The government can send notices, health recommendations, and record people's feedback to take necessary initiatives to prevent or control the spread of COVID-19.

7) Treatment services - Around half (48%) of the applications provide treatment services or treatment-related information. The available functionalities to achieve this objective are: search functionality to find nearby medical facilitates with COVID-19 testing capacity and pharmacies; provide information about the nearby hospitals, hospital maps, and travelling information to reach the nearby hospitals. Some applications provide information related to COVID-19 symptoms so that in case of suspected infection, users may conduct the self-assessment. Some other apps recommend suspected patients to contact the nearby service centers or hospitals based on their geographical location and physical symptoms.

8) Raise awareness - Similar to the current status objective, most of the applications (i.e., 17 out of 25 applications) make people aware of the pandemic by providing the recent updates (corona news and corona status). In addition, some applications provide useful web links related to the COVID-19, information about when, how, and why to stay at self-isolation. A few apps provide declarations about new suspected cases, death, recovery, etc. from a local (a specific country) perspective and the push notification for any important/emergency alert message and news about suspected community/regions.

9) Improve mental health - A total of seven (28%) applications focused on users' mental health which is very crucial for each individual during the pandemic. These applications improve motivation and self-confidence of the users through community building, sharing personal experiences with other people, and by providing the positive COVID-19 news like recovery updates of the patients and recent/new initiatives taken by the government in order to prevent and control the spread of COVID-19.

### B. APP CONTEXT, LANGUAGE AND USER RATINGS

The summary of the evaluated apps as presented in Table 1, shows that among the existing applications, 4 apps (16%) were developed exclusively in iOS platform, two apps (8%) were developed in Windows platform, 12 apps (48%) were developed in Android platform and the remaining apps were developed in both iOS and Android platforms. All the apps were freely available. Each application mentioned about its privacy policy. More than half of the applications (n=13) were developed for a specific country context that includes India, Vietnam, Brazil, France, Bangladesh, UK, Canada, Spain, Australia, and Greece. Among the apps, twelve were developed in local language and one app COVA Punjab was in English (see Table 1).

### C. MINING THE APP REVIEW

A total of ten information systems design characteristics were found on the mobile applications developed targeting COVID-19 through mining the app reviews. Both positive and negative experiences were found in each category of design characteristics. The design characteristics codes and the examples of users' comments are presented with the + (positive) and - (negative) signs in Table 2. The design characteristics related factors (see Table 2) are discussed below.

*Reliability* refers to the dependability of the service [36]–[38] such that it enables users to receive authentic and updated information, and make sure that the app contents don't encourage panic but promote awareness. *Performance* captures the predictability and functionality dimensions [36]. Therefore, performance relates to issues such as well-execution of the app functionalities, help in slowing down the spread of the virus, and failure to access some of the app functionalities properly. *Usefulness* and *Supportive* are concerned with the extent to which an application helps the users to achieve their purposes such as raise awareness about the coronavirus, properly keep track of the updated situation, support for self-caring, and maintain social distancing. We note that the concept of perceived usefulness has been widely used in prior literature and has been suggested as one of the main predictors of adopting and using a service [39]–[41]. *Security* and *Privacy* relate to the users' concern on digital privacy, accessing the application anonymously, unnecessary resource access permission, and asking for personal information or phone number during sign-up. Both security and privacy have been suggested important design considerations in prior literature [42]. *Flexibility* refers to the way the system adapts to the changing demands of the users [36], [37]. Therefore, in our context flexibility is about allowing the users to post without world limit, avoiding rigid information (e.g., phone number) input (i.e., in a specific data format), and allowing options to hide the posts and setting up reminders. Concerns related to receiving immediate feedback, activation code or OTP, and fast response/load are related to the *responsiveness* factor of information systems characteristics. The responsiveness characteristics can be captured through the timeliness and speed characteristics that have been discussed in prior literature [36]. *Ease-of-use and usable* refers to the design aspect of user screens in terms of ease of use, visual attractiveness, user-friendliness, and convenience in delivering services [39]. Therefore, these information systems design characteristics are concerned with the easy and convenient way to access information, use of intuitive navigation buttons and visual representations, and proper organization of the app contents (information architecture). Finally, the *culturally sensitive* factor refers to developing contextual (in terms of country and language) applications for more effective and efficient use of the app functionalities by the local people. We note that language has been emphasized by Wixom and Todd [37], whereas Zakaria et al. [43] emphasized on designing

**IEEE** Access·

**TABLE 2.** Design characteristics, codes and a set of examples of user comments

| Ser | Design Characteristics | Codes | Examples of User Comments |
|---|---|---|---|
| 1 | Reliability | + Authentic information<br>+ Respectful in publishing info<br>+ Regular update<br>- Spread false info<br>- Discuss unrelated subject<br>- Encourage panic | (+) *"Information being shared with users is accurate"*<br>(-)*"Instead of encouraging awareness, it encourages panic"*<br>(-)*"Spreading of false information, and discussing unrelated subjects"* |
| 2 | Performance | + Well-functioning<br>+ Slow down the virus spreading<br>- Incompetence<br>- Unable to login<br>- Important features missing | (+) *"As soon as I see a new cases or deaths in the news, I check the app, and it has the updated numbers. Good maps, you can see the confirmed cases, deaths and more."*<br>(+) *"Can slow but not stop the spread of virus"*<br>(+) *"The app even send me a phone call just to inform my physical & mental health so that I could ask any way for a diagnose or doctor service. Really appreciable service."*<br>(-) *"Leaves a lot of rooms for growth and improvement"*<br>(-) *Tried to login from last 30mins but it's only buffering.* |
| 3 | Useful | + Helpful app<br>+ Important app<br>+ Best app<br>+ Wonderful app<br>+ Very useful app<br>+ Good use<br>- Worthless app<br>- Worst app | (+) *"Thanks for inaugurated such type of handy App. We'll be benefited by using this"*<br>(+) *"Great app is being very helpful for the purpose it was made"*<br>(+) *"A PERFECT app for updates about the pandemic"*<br>(+) *"Good app to keep track of the situation."*<br>(+) *Such a wonderful app.. provides information in easy and convenient method.. cool navigation links.. well done team for doing such noble cause.*<br>(-) *"This app is the worst"*<br>(-) *"Worst app with no necessary function. Data given in app is not updated regularly"* |
| 4 | Supportive | + Raise awareness<br>+ Support health ministry<br>+ Support self-care<br>+ Help to maintain social distance<br>+ Updated with the situation<br>+ Provide important info<br>- Worthless app<br>- Contain very little information | (+)*"Raising calm awareness is the objective and seeing cases over time"*<br>(+) *"Many thanks for coordinating to popularize information to people quickly and promptly."*<br>(-) *"Trivial. . . Contains very little info compared to the J. Hopkins website for instance"* |
| 5 | Secure | + Don't ask for personal info<br>- Fair about digital privacy<br>- Ask unnecessary resource permission<br>- Ask unnecessary personal information | (-)*"Takes too much resource permission to edit and read"*<br>(-) *"I am concerned about digital privacy and this is just not okay, especially in this crucial time when information is crucial"* |
| 6 | Privacy | + Hide post (if needed)<br>+ No signup required<br>+ Anonymously accessible<br>- Require phone number to sign up<br>- Ask unnecessary personal information | (+) *"There is no sign up required, and you stay behind a generic"*<br>(+) *"There is no sign up required, and you stay behind a generic name and avatar (Arlo, Moki, Paskel, etc)"*<br>(-) *"I did not like that you have to sign up with your phone number before you can enter the app"* |
| 7 | Flexibility | + No word limit to post<br>+ Hide post (if needed)<br>+ Setup reminder for warning<br>- Not accept a valid phone number | (+) *"There is no word limit for your posts"*<br>(+) *"You can hide posts if you feel uncomfortable or report if it's bad"*<br>(-) *"Repeatedly ask for the valid phone number though the number is valid according to users opinion"* |
| 8 | Responsiveness | + Receive feedback<br>+ React immediately<br>- OTP is not received<br>- Activation code is not received<br>- Keep loading/ slow response<br>- Terribly slow | (+*"Always receive feedback & reactions"*<br>(-) *"I haven't received any OTP yet while installing it"* |
| 9 | Usable/<br>easy-to-use | + Easy and convenient to access info<br>+ Intuitive navigation button/link<br>+ Understandable visual representation<br>- Require help to launch<br>- Poor organization of contents | (+)*"Provides information in easy and convenient way. . . cool navigation links"*<br>(-)*"I think this app would be more helpful if stats were organized by country, rather than by region. . . such anomalies would be easier to detect.Thank you."*<br>(-) *"Get the OTP code, but don't know where to click so it will continue"* |
| 10 | Culturally<br>sensitive | + Helpful for Bangladeshi<br>+ Help Brazilian<br>+ Useful for Panjab people<br>+ Useful for Vietnamese<br>- Don't understand the language | (+) *"In case of testing corona virus, some may be confused whether to go to take treatment or not. In this case, this app is very much helpful for the people of Bangladesh."*<br>(+) *"Good app for tracking punjab COVID Stats alongwith curfew pass, labour registration, report mass gathering, foreign travelers. It's good that govt is adding more public services to this app, Weldone"*<br>(-) *"Not in English. So, can't read"* |

and implementing culturally sensitive IT applications considering privacy and cultural values.

Each of the above information systems design characteristics related factors are mentioned both positively and negatively in the app review. For example, in the case of *reliability* factor (see Table 2), users' concerns related to providing authentic information, respectfulness in publishing information, and updating information in a regular basis were positive user experiences; while user comments related to providing insufficient information, spreading false information, discussing unrelated subjects and creating panic were negative user experiences.

## IV. DISCUSSION

The review found that only a limited number of mobile applications have been developed. However, we note that more application development efforts will eventually come in the future. Our review revealed the main purposes of developing the mobile applications and the functionalities to achieve these objectives for the prevention, mitigation, and containment of COVID-19. The review study also explored the key factors or concerns that affect the end-user experience. Our results indicate that users prefer applications with higher reliability, performance, responsiveness, supportive, ease-of-use, usefulness, security, privacy, and flexibility. We also observed that culturally sensitive applications are needed. COVID-19 pandemic creates a crucial situation in the affected countries.

### A. IMPLICATIONS

This research has implications for future research and practice. The outcomes of this review will greatly contribute to the health institutes, health workers, practitioners, and the governments of the COVID-19 affected countries. Our paper summarizes the currently developed apps to raise awareness about the existing mobile applications, their functionalities, and design characteristics. The findings can help to take necessary initiatives in developing new applications or updating the existing applications to receive maximum benefits out of these applications during the pandemic. The review study also can be considered as a requirement elicitation study. The app developers may consider the revealed objectives and functionalities as the user requirements. For example, if the professionals or governments want to develop a mobile application to ensure population awareness about the COVID-19 pandemic, they may incorporate the services associated with the 'raise awareness' objective. In contrast, if a government would like to develop a new application to provide all possible services related to COVID-19, it may consider the nine objectives and the entire set of functionalities revealed from this study as shown in Figure 2.

This review also helps to understand the key information system design characteristics that need to be addressed to develop such applications. The revealed design characteristics can be considered as the design recommendations to the practitioners for developing and evaluating such appli-cations. The design recommendations based on the revealed design characteristics are as follows. The applications should provide quick and accurate responses. The apps need to be usable and useful. The applications should provide authentic information in order to be reliable. The applications should also perform well with the required functionalities. This will also ensure that the applications are useful and supportive with respect to the user requirements. The security and privacy issues should be addressed properly since these aspects are critical to many users during the vulnerable time. Responsive and flexible app design and development need to be ensured. The application should be easy to use in order to ensure that different types of users can use it. Finally, the application should be culturally sensitive, when possible; since users may prefer to have a contextual (local) app in their own language.

In sum, the findings of our study would be a great source of inspiration for governments to take necessary initiatives to develop new applications and promote the adoption of existing/new mobile application(s). Finally, for app development companies, these outcomes provide an indication for developing new and innovative mobile apps targeting the affected countries.

### B. LIMITATIONS

The study presented in this paper has a few limitations that are important to acknowledge. Firstly, the criteria (search strings/keywords) chosen for selecting the relevant apps for this study may not cover all the available applications, especially the applications that are named in a local language. Secondly, this study does not include the relevant apps that are developed or became available in the app stores after April 30, 2020. Thirdly, the review data was analyzed through the qualitative approach using the affinity diagram and *noticing-collecting-thinking*. Qualitative analysis is subjective and partly depends on analyzers' skills, expertise, and knowledge. Therefore there might be some flaws in the data extraction, coding, and clustering process. However, in order to alleviate these limitations, researchers meticulously conducted checks for apps selection and data analysis, and made adjustments where necessary through discussions. Fourth, in this study, we did not investigate which features are essential and which are nice-to-have. Thus, a future study may explore or classify the essential and nice-to-have features to achieve the identified objectives. Finally, the online review data collected from the app stores may contain biases. Therefore, future research can use multiple methods to collect and analyze data to investigate the validity of our findings.

.

**IEEE** *Access*

# APPENDIX A  DATASET

**TABLE 3.** Mapping between features and mobile applications

| Ser. | Features | A1 | A2 | A3 | A4 | A5 | A6 | A7 | A8 | A9 | A10 | A11 | A12 | A13 | A14 | A15 | A16 | A17 | A18 | A19 | A20 | A21 | A22 | A23 | A24 | A25 | Freq |
|---|---|---|---|---|---|---|---|---|---|---|---|---|---|---|---|---|---|---|---|---|---|---|---|---|---|---|---|
| 1 | Health monitor | X | | | | | | | X | X | | | X | | | | | X | X | X | X | X | X | X | | | 11 |
| 2 | Video consultation | X | | | | | | | | | X | | | | | | | | | | | | | | | | 2 |
| 3 | Self-reporting | | X | | | | | | | X | | | X | | | | | | | | | | | | | | 3 |
| 4 | Corona Status | | X | X | X | | | X | | X | | X | X | | X | X | X | | X | | | X | | | X | X | 14 |
| 5 | Chatting | | X | | | | | | | | | | | X | | | | | | | | | | | | | 2 |
| 6 | Corona news | | X | | | X | X | X | | | | | X | | | | | | X | | | | | | X | X | 8 |
| 7 | Push notification | | | X | | X | | | | | | X | | | | | | | | | | | | | | X | 4 |
| 8 | State/data visualization | | | | X | X | | | | | | X | | | | X | X | | | | | | X | | X | | 7 |
| 9 | COVID Symptoms | | | | | | | | X | | | | | | | | X | | | | | | | X | X | X | 5 |
| 10 | Prevention guidelines | | | | | | | | X | X | | X | X | | | X | | X | | X | X | | | X | X | X | 11 |
| 11 | Nearby hospital map | | | | | | | | X | | X | X | X | | | | | | | | | | | | | X | 5 |
| 12 | Self-assessment | | | | | | | | X | | | | X | | | | | X | X | | | | | | | X | 5 |
| 13 | Recommend hospital | | | | | | | | X | | | | | | | | | | | | | | | | | | 1 |
| 14 | Call Emergency | | | | | | | X | | X | | | X | | | | | | X | | | | | | | | 4 |
| 15 | Provide web link | | | | | | | X | | | | | | | | | X | | | | | | | | | X | 3 |
| 16 | Service info in crises | | | | | | | | X | | | | | | | | | | | | | | | | | | 1 |
| 17 | Self-isolation info | | | | | | | | | X | | | | | | | | | | | | | | | | | 1 |
| 18 | Virtual assistant | | | | | | | | | | X | | | | | | | | | | | X | | X | | | 3 |
| 19 | User Feedback to Govt | | | | | | | | | | X | | | | | | | | | | | | | | | | 1 |
| 20 | Report suspected case | | | | | | | | | | | | | X | | | | | | | | | | | | | 1 |
| 21 | Motivational info | | | | | | | | | | | | | | X | | | | | | | | | | | | 1 |
| 22 | self-confidence | | | | | | | | | | | | | | X | | | | | | | | | | | | 1 |
| 23 | Medical Notice | | | | | | | | | | | | | X | | | | | | | | X | | | | | 2 |
| 24 | Call support | | | | | | | | | X | | | | | | | | | | | | | | | | | 1 |
| 25 | Corona hospitals | | | | | | | | | | X | | | | | X | | | | | | | | | X | | 3 |
| 26 | Warning Notification | X | | | | | | | | | | | | | | | | | | | | | | | | | 1 |
| A1, A2, A3,.....,A25 : The number of the mobile applications as shown in Table 1. | | | | | | | | | | | | | | | | | | | | | | | | | | |

**TABLE 4.** Mapping between objectives and the mobile applications

| Ser. | Objectives | A1 | A2 | A3 | A4 | A5 | A6 | A7 | A8 | A9 | A10 | A11 | A12 | A13 | A14 | A15 | A16 | A17 | A18 | A19 | A20 | A21 | A22 | A23 | A24 | A25 | Freq | Percent |
|---|---|---|---|---|---|---|---|---|---|---|---|---|---|---|---|---|---|---|---|---|---|---|---|---|---|---|---|---|
| 1 | Remote assistance | X | X | X | | | X | X | | X | | X | X | X | | | | | X | X | | X | | X | | X | 14 | 56 |
| 2 | Monitor patients | X | | | | | | | X | X | X | | X | | | | | X | X | X | X | X | X | | | X | 12 | 48 |
| 3 | Current status | | X | X | X | X | X | X | X | X | X | X | X | | X | X | X | | X | | | | | X | X | X | 18 | 72 |
| 4 | Raise awareness | | X | X | X | X | X | X | X | | X | X | | | X | X | X | | X | X | | | | X | X | X | 17 | 68 |
| 5 | Communication support | X | X | | | | | | X | X | X | X | | X | | | | | X | | | | | | | | 8 | 32 |
| 6 | Prevent COVID-19 | | | | X | X | X | | X | X | X | X | | X | | | X | | X | X | | X | | X | X | X | 15 | 60 |
| 7 | Improve mental health | | | | X | X | X | | | | | | | X | X | | | | | | | | | | X | X | 7 | 28 |
| 8 | Control COVID-19 | | | | X | X | X | | | X | X | | X | | | | X | | X | | | | | | X | X | 10 | 40 |
| 9 | Treatment services/information | | | | X | X | | | | X | X | X | | | | | X | | X | X | | X | | X | X | X | 12 | 48 |
| A1, A2, A3,.....,A25 : The number of the mobile applications as shown in Table 1 | | | | | | | | | | | | | | | | | | | | | | | | | | | |

**IEEE** Access·

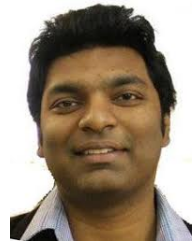

**A.K.M. NAJMUL ISLAM** is an Adjunct Professor at Tampere University, Finland. He is a Scientist at LUT University, Finland. He also works as University Research Fellow at the Department of Future Technologies, University of Turku, Finland. Dr. Islam holds a PhD (Information Systems) from the University of Turku, Finland and an M.Sc. (Eng.) from Tampere University of Technology, Finland. He has 95+ publications. His research focuses on Human Centered Computing. His research has been published in top outlets such as IEEE Access, European Journal of Information Systems, Information Systems Journal, Journal of Strategic Information Systems, Computers in Industry, Technological Forecasting and Social Change, Computers in Human Behavior, Internet Research, Computers & Education, Journal of Medical Internet Research, Information Technology & People, Telematics & Informatics, Journal of Retailing and Consumer Research, Communications of the AIS, Journal of Information Systems Education, AIS Transaction on Human-Computer Interaction, and Behaviour & Information Technology, among others.

• • •

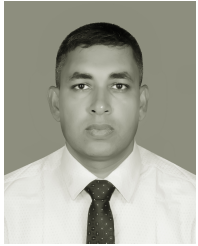

**MUHAMMAD NAZRUL ISLAM** received the B.Sc. degree in computer science and information technology from the Islamic University of Technology, Bangladesh, in 2002, the M.Sc. degree in computer engineering from Politecnico di Milano, Italy, in 2007, and the Ph.D. degree in information systems from Åbo Akademi University, Finland, in 2014. He is currently an Associate Professor with the Department of Computer Science and Engineering, Military Institute of Science and Technology (MIST), Mirpur Cantonment, Dhaka, Bangladesh. Before joining MIST, he was working as a Visiting Teaching Fellow with Uppsala University, Sweden and as a Postdoctoral Research Fellow with Åbo Akademi University, Finland. He was also a Lecturer and an Assistant Professor with the Department of Computer Science and Engineering, Khulna University of Engineering and Technology (KUET), Bangladesh, from 2003 to 2012. His research areas include but not limited to human–computer interaction (HCI), humanitarian technology, health informatics, military information systems, information systems usability, and computer semiotics. He is the author of more than 80 peer-reviewed publications in International journals and conferences. He is a member of The Institution of Engineers, Bangladesh (IEB).

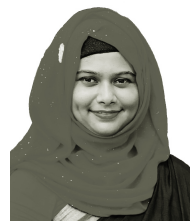

**IYOLITA ISLAM** received the B.Sc. degree in computer science and engineering from the Military Institute of Science and Technology (MIST), Dhaka, Bangladesh, in 2017, where she is currently pursuing the M.Sc. degree in computer science and engineering. She is serving as a Lecturer with the Computer Science and Engineering Department, MIST, since 2018. She is the author of several conference papers. Her research interests include human–computer interaction, blockchain technology, health informatics, and industry 4.0. She was awarded the Best Paper Award in International Conference on Sustainable Technologies for Industry 4.0 (STI 2019). She is an Associate Member of The Institution of Engineers, Bangladesh (IEB).

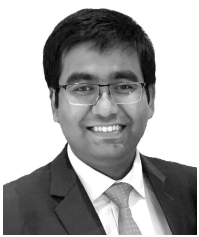

**KAZI MD. MUNIM** received the B.Sc degree in computer science and engineering from the Military Institute of Science and Technology (MIST), Dhaka, Bangladesh, in 2017, where he is currently pursuing the M.Sc. degree. He is currently a Software Quality Assurance Engineer by Profession. His research interests include human–computer interaction, blockchain, and industry 4.0. He is the author of several conference papers. He was awarded the Best Paper Award in International Conference on Sustainable Technologies for Industry 4.0 (STI 2019).